\documentstyle[12pt]{article}
\renewcommand{\theequation}{\arabic{section}.\arabic{equation}}
\newcommand{\thbyth}[9]
{ \left[ \begin{array}{ccc}
        {#1}   &  {#2}  &  {#3} \\
        {#4}   &  {#5}  &  {#6} \\
        {#7}   &  {#8}  &  {#9} \end{array}  \right] }
\newcommand{\twbytw}[4]
{ \left[ \begin{array}{cc}
        {#1}   &  {#2} \\
        {#3}   &  {#4} \end{array}  \right] }
\newcommand{\der}[2]{\frac{{\rm d}{#1}}{ {\rm d} {#2}} }
\newcommand{\pdr}[2]{ \frac{\partial  {#1}}{\partial {#2}  }}
\newcommand{\pddr}[3]{ \frac{\partial ^2 {#1}}{\partial {#2} \partial {#3} }}
\newcommand{\vc}{\vec{c}}
\newcommand{\vv}{\vec{v}}
\newcommand{\fot}[1]{$^{\ref{#1}}$}
\newcommand{\fow}[2]{$^{ \ref{#1}-\ref{#2} }$}
\newcommand{\refl}[1]{[\ref{#1}]}

\newtheorem{Proposition}{Proposition}
\begin{document}
\baselineskip .333in
\title{{\bf Complex Trkalian Fields and Solutions
to Euler's Equations for the Ideal Fluid}}
\author{{\sf P. R. Baldwin }\\
       {\sf G. M. Townsend }\\
   {\small\it Department of Physics } \\
   {\small\it The University of Akron} \\
   {\small\it Akron, OH 44325-4001}}
\maketitle

\vspace{-5ex}
\begin{center} {\bf Abstract} \end{center}
We consider solutions to the complex Trkalian
equation,~$ \vec{\nabla} \times \vc = \vc ,$
where~$\vc$ is a 3 component vector function with
each component in the complex field, and may be expressed
in the form~$  \vc = e^{ig} \vec{\nabla} F, $ with~$g$ real
and~$F$ complex.
We find, there are precisely two classes of solutions;
one where~$g$ is a Cartesian variable and one
where~$g$ is the spherical radial coordinate.
We consider these flows to be the simplest of
all exact 3-d solutions to the Euler's equation
for the ideal incompressible fluid.
Pictures are presented.

The novel approach we use in solving for these classes
of solutions to these 3-dimensional vector pdes
involves differential geometric techniques:
one may employ the method to  generate solutions
to other classes of vector pdes.

\vspace{2ex}
\noindent
Keywords:\hspace{6ex} {\em Euler Equations,
Fluid Dynamics, Beltrami Fields}\\
Pacs Numbers:\hspace{1ex} 41.20.-q, 47.20.-k, 47.90.+a

%%%%%%%%      SECTION 1 %%%%%%%%%%%%%%%%%%%%%%%%%%%%%%%%%%%%%%
%%%%%%%%                %%%%%%%%%%%%%%%%%%%%%%%%%%%%%%%%%%%%%%
%%%%%%%%                %%%%%%%%%%%%%%%%%%%%%%%%%%%%%%%%%%%%%%
\section{Introduction}
\setcounter{equation}{0}
\hspace{4ex}
The Beltrami equation
\begin{equation}  \vv \times (\vec{\nabla} \times \vv)=0  \end{equation}
has received enormous attention in recent years.
Reviews$^{ \ref{Wan1}- \ref{Wan3}}$ note the
prominent role that Beltrami fields
play in the theory of exact,
closed form solutions to the Euler's and
Navier Stokes equations and their relations
to the electromagnetic wave equations.
Moreover, Beltrami fields are related to minimum energy
plasma fields and have therefore garnered much attention
from the MHD community.$^{ \ref{MonDahPhiThe}- \ref{Dri}}$
In this paper, we would like to undertake the
first steps in a systematic classification of
these vector fields, using a novel
differential geometric technique which we discss below.

Any 3 component vector function,~$\vv$,
in~${\cal R}^3$ may be written as the real part of
a three component complex vector field,
of the following form:
\begin{equation}  \vc = e^{ig} \vec{\nabla} F. \end{equation}
where~$g$ is a real function and~$F$ is a function
from~${\cal R}^3$ to~${\cal C}^3$.
Trkalian fields\fot{Trk} are real vector fields which solve
\begin{equation} \vec{\nabla} \times \vv = \vv , \end{equation}
and therefore satisfy 1.1.
We solve the complex version of the Trkalian equation:
\begin{equation} \vec{\nabla} \times \vc = \vc , \end{equation}
where~$\vc$ is also a vector of the form of Eq. (1.2).
Notice that both the real and imaginary parts of~$\vc$
must solve the Trkalian equation, which is a linear equation.
{}From solutions to the Trkalian equation, one
straightforwardly may derive (non-zero helicity)  three dimensional
(in a topological sense: see \refl{PRB1})
time-independent solutions to Euler's equations for
the incompressible ideal fluid\fot{NemPri} and time dependent
solutions to the incompressible Navier Stokes equations.\fot{Mar2}
Therefore we consider the equations (1.2) and (1.4)
to be the simplest possible three-dimensional
hydrodynamical equation.
Although we do not discuss the stability of
these solutions, a discussion of boundary conditions
is given in section 4.

We completely categorize all solutions to the simulataneous
set of equations given by (1.2) and (1.4).
The two classes of solutions to this problem are A.) where
the function~$g$ in 1.4 is identical to a Cartesian variable
(for example:~$z$) and
the function~$F$ is a complex analytical function of the
other two Cartesian variables (see 3.12),
and B.) where~$g$ is the spherical radial coordinate~$r$ and
the function~$F$ is an analytical function of a certain
combination of the spherical angular variables
(see 3.19).
In both cases~$F$ and also~$g$ each solve Laplace's equation.
Theses solutions are closely related to the possible
forms of linearly polarized TEM waves.\fot{PRBTEM}

Our method of solution is very particular and
and we will employ it in further work. Finding the solution
to equations (1.2), (1.4) is a first step in an attempt
to solve three dimensional hydrodynamical problems using
potential methods. The procedure is as follows.
The transformation from the three Cartesian variables
variables to the functions~$g,F, \bar{F}$ leads
naturally to a metric function~$q_{ij}$
\begin{equation} {\rm d}s^2 =  {\rm d}x_1^2 + {\rm d}x_2^2 +  {\rm d}x_3^2 =
q_{ij} {\rm d}y^i {\rm d}y^j . \end{equation}
Where we have used~$y_1 = g, \; y_2=F, \; y_3 = \bar{F} $.
First, the partial differential Eq. 1.4 may be
interpreted as constraints on the metric tensor:
the appropriate relations among the elements
of the metric tensor~$q_{ij}$ are found.
Next with this metric one may calculate
the Ricci tensor.
In dimensions fewer than 4, the Riemannian curvature
tensor and the Ricci tensor are linearly
related one to another (see Landau \& Lifshitz\fot{LanLif2},
for example).
Indeed, the vanishing of the Ricci tensor is both a
necessary and sufficient condition that
the coordinate system~$g, \; F, \; \bar{F} $ be a
diffeomorphism of the Cartesian~$x_1, x_2, x_3$ system.
It is interesting that only in the very last step of the
procedure, that one makes the correspondence of the~$g, \; F, \; \bar{F}$
system with the Cartesian system. That is, the original
vector pde is solved with respect to the natural coordinates
of the problem, which are not in general the Cartesian variables.

In section 2, we show how the solutions to 1.1 are related
to solutions to Euler's equations.
In section 3, we completely characterize the solution
to 1.1 by the method outlined above.
Since the method is novel,
we show what the method might look like as applied in
solving two-dimensional incompressible, potential flow
problems in Appendix B. Of course, in the latter case, the solutions
may be derived straightforwardly by other
well known means; the discussion is included since it
isolates the method particularily well
(for some already familiar problems).
The reader is strongly encouraged to read
Appendix B, before beginning section 3.

Section 4 discusses what sort of boundary conditions
the complex Trkalian fields may satisfy.
The type A fields do not satisfy
natural boundary conditions, however the
type B (singular) fields may have no normal component
to a bounding surface.

Section 5 discusses how the work presented here
meshes with
Bj\mbox{\o}rgum\fot{Bjo} and  Bj\mbox{\o}rgum and Godal's\fot{BjoGod}
work on Beltrami fields.
Section 6 is a discussion and conclusion.
In ongoing work, we discuss solutions to
other vector pde's: including linearly polarized
transverse electric field\fot{PRBTE} (TE)  and TEM\fot{PRBTEM}
solutions to the the electromagnetic wave equations.
The main purpose of this article is
to introduce a potential method capable of
constructing truly three dimensional vector
field solutions to pde of mathematical physics.

A glossary of various terms is appended
at the end of the manuscript.
%Also is included Appendix A concerning
%zero-helicity, complex vector fields.

%%%%%%%%      SECTION 2 %%%%%%%%%%%%%%%%%%%%%%%%%%%%%%%%%%%%%%
%%%%%%%%                %%%%%%%%%%%%%%%%%%%%%%%%%%%%%%%%%%%%%%
%%%%%%%%                %%%%%%%%%%%%%%%%%%%%%%%%%%%%%%%%%%%%%%
\section{Hydrodynamics and The Complex Trkalian Field}
\setcounter{equation}{0}
\hspace{4ex}
Trkalian fields yield solutions to Euler's equations
for an incompressible inviscid fluid.
These latter equations may be written:
\begin{equation} \pdr{\vv}{t} + (\vv \cdot \vec{\nabla}) \vv =
    - \frac{\vec{\nabla} p}{\rho}, \end{equation}
where the density~$\rho = {\rm constant};$
and~$\; \vec{\nabla} \cdot \vv = 0.$
A pressure function may be found (at least locally), %\fot{exception}
if and only if the vorticity equation is satisfied:
\begin{equation} \pdr{\vec{\xi}}{t} + \vec{\nabla} \times
\left( \vec{\xi} \times \vv \right) = 0 , \end{equation}
where~$ \vec{\xi} = \nabla \times \vv $.
Thus any time independent vector field, satisfying
$\vec{\xi} \times \vv = 0 $ is a
solution to Euler's incompressible
equations.$^{\ref{Cro}-\ref{HicGueWas}}$
We shall call these flows
Beltrami flows (as is in keeping with Bj\mbox{\o}rgum).
A subset of these flows are the Trkalian fields,
satisfying~$ \vec{\xi} = \lambda \vv , $ where~$\lambda$
is a constant, which under suitable scaling
and inversion we can always take to be unity.

Consider therefore the Trkalian flows which solve the differential equation,
 \begin{equation}  \vec{\xi} = \vec{\nabla} \times \vv = \vv. \end{equation}
Trkalian fields are divergenceless and obey a linear
superposition principle.
Each component of the vector field satisfies the Helmholtz
eigenvalue equation,
\begin{equation} \triangle v_i  + v_i =0 .\end{equation}
In the literature therefore, Trkalian fields
are often represented as a sum over Fourier modes,
using wavevectors with unit
norm:
\begin{equation}   (\vec{a}_{\vec{k}} +
  i \vec{k} \times \vec{a}_{\vec{k}} )
   e^{i \vec{k} \cdot \vec{r}} ,  \end{equation}
where~$\vec{k} \cdot \vec{a}_{\vec{k}} =0$
and~$ \vec{k}\cdot \vec{k} = 1$.
It is our contention that the Fourier type
representation (2.5) is not necessarily the most
useful representation of Trkalian fields.

Time-independent Trkalian fields (2.3) clearly
solve the vorticity Eq. (2.2)
for the incompressible Euler's equations.
Trkalian fields multiplied by the time decaying
factor~$e^{-\nu t}$ are solutions to the incompressible Navier Stokes
equations (Eq. 2.1 with a term~$\nu \triangle \vv$
included on the RHS, where~$\nu$ represents the viscosity: see
\refl{Mar2}.)

For this paper we are interested in those vector fields
which solve the Trkalian equation 2.3, and have a
``topologically dual field,''~$\vec{w}$, which also solves
the Trkalian equation. The relations between vector fields
and their topological duals have recently been investigated
by one of the authors;\fot{PRB2}
some results are collected in Appendix A.
In summary, a vector field,~$\vv$, and a dual field,~$\vec{w}$,
are related in such a way that
\begin{equation}  \vec{c} = \vv + i \vec{w} = e^{ig} \vec{\nabla} ( F_1 + i
F_2), \end{equation}
for some~$g, \;F$~and~$\bar{F}$,
where~$g, F_1, F_2$ are all real functions.
As pointed out in the Appendix {\bf any real vector field may be
represented as the real part of a complex vector field, written
as 2.6}.
That is~$\vec{c}$ (as well as~$\vec{w}$)  may be expressed with
precisely the same three functions with which one
may express~$\vv$.
Among other properties, a vector field is always
orthogonal to the curl of a dual field, and a field
and a dual field both have precisely the same value of the
helicity~$H= \vv \cdot \vec{\nabla} \times \vv=  \vec{w} \cdot \vec{\nabla}
\times \vec{w} $.
A given positive helicity vector field does not in general
have a unique dual field (although a suitable gauge
fixing condition could probably be defined.)

We call the simplest possible solution to (1.1) and (1.2)
``the simple helical field'':
\begin{equation}  \vv = ( \sin z, \cos z, 0), \end{equation}
with
 \begin{eqnarray} \vec{w} &=& (- \cos z, \sin z, 0),  \\
     \vc = \vv + i \vec{w} &=& e^{iz}( -i, 1, 0)
        = e^{iz}\vec{\nabla} (x+iy)\cdot(-i) . \end{eqnarray}
Eq 2.9 is obviously of the form of (1.1).
Notice that~$\vv$ and~$\vec{w}$ are orthogonal
and both have unit helicity (implying~$\vc \cdot \vc =0$ where
dot is the usual scalar product).
For the Trkalian vector fields,
we may very roughly think of a dual field
as a rotation of each vector around the screw
axis (in this case the~$z$-axis), where the screw axis is
along the~$\vec{\nabla} g$ direction.

%%%%%%%%      SECTION 3 %%%%%%%%%%%%%%%%%%%%%%%%%%%%%%%%%%%%%%
%%%%%%%%                %%%%%%%%%%%%%%%%%%%%%%%%%%%%%%%%%%%%%%
%%%%%%%%                %%%%%%%%%%%%%%%%%%%%%%%%%%%%%%%%%%%%%%
\section{Solution to the Complex Trkalian Equation}
\setcounter{equation}{0}
\hspace{4ex}
The reader is urged to read Appendix B so that the logic used
in our approach will be completely clear.
We begin with the complex Trkalian equation:
\begin{equation} \vec{\nabla} \times \vc = \vc . \end{equation}
Suppose in addition that
\begin{equation} \vec{\nabla} \times \vc = i \vec{\nabla} g \times \vc ,
\end{equation}
where~$g$ is some real function.
Eq. 3.2 implies that~$ \vc = e^{ig} \vec{\nabla} F ,$
for some complex valued~$F$.
That is,~$\vec{c}$ is a zero-helicity field of a very
special type.
(In general,~$ \vc = G \vec{\nabla} F ,$ where~$F,G$ are complex,
yields ~$\vec{c} \cdot \vec{\nabla} \times \vec{c} = 0$.)
Substituting the relation,~$\vc = e^{ig} \vec{\nabla} F$, into 3.1 yields~
\begin{eqnarray} \vec{\nabla} g \times \vec{\nabla} F &=& -i \vec{\nabla} F ,
\\
      \vec{\nabla} g \times \vec{\nabla} \bar{F} &=&  + i \vec{\nabla} \bar{F}
,
\end{eqnarray}
where~$\bar{F}$ is the complex conjugate of~$F$.
We wish to find the components of the
tensor,~$q_{ij} \equiv  \pdr{x^p}{y^i} \pdr{x^p}{y^j}$,
where we
define~$y_1 \equiv g, \; y_2 \equiv F, \; y_3 \equiv \bar{F} $.
First we note that this tensor and the
tensor~$q^{lk} \equiv \pdr{y^l}{x^p} \pdr{y^k}{x^p} $
are matrix inverses. Next from (3.3), for example,
one immediately notes that~$\vec{\nabla} F \cdot \vec{\nabla} F =0$.
Thus~$q^{22}=0$. Similarily we find~$q^{33} =0,
\; q^{13}=q^{12}=0$ and~$q^{11} =1. $
The last result comes from taking the inner product
of (3.3) and (3.4) and simplifying.
Finally define~$H' \equiv  \vec{\nabla} y^1 \cdot \vec{\nabla} y^2
\times \vec{\nabla} y^3 $, which turns out to be a pure complex
entity.
As is always true, we have~$  \det q^{ab} = (H')^2  $.
Putting these together, we may write
\begin{equation}  q^{ab} =
      \thbyth{1}{0}{0}{0}{0}{iH'}{0}{iH'}{0} ,\end{equation}
with matrix inverse
\begin{equation}  q_{ab} =
   \thbyth{1}{0}{0}{0}{0}{\frac{1}{iH'}}{0}{\frac{1}{iH'}}{0} .\end{equation}
1. This is equivalent to
\begin{equation} {\rm d}s^2 =  {\rm d}g^2 +
      2 e^{2J} {\rm d}F  {\rm d}\bar{F} , \end{equation}
 if we
define~$e^{2J}$ (as yet
undetermined) to be~$\frac{1}{iH'}$.
\underline{ The non-vanishing components of the}
\underline{metric tensor have been identified.}
This completes the first step of the procedure.

2. At this point we have just calculated
the metric tensor.
We must ensure however that the various coordinate
relations encoded into the metric tensor are
consistent with a flat (Euclidean)
three-dimensional space.
We begin with the line element 3.6 (or 3.7)
and calculate the Ricci tensor. First
\begin{equation}  \Gamma^i_{kl}
             \equiv  \frac{1}{2} q^{im}
   \left( \pdr{q_{mk}}{y_l} +  \pdr{q_{ml}}{y_k}
          -  \pdr{q_{kl}}{y_m} \right). \end{equation}
Then
\begin{eqnarray} R_{bc} &\equiv& \Gamma^a_{bc,_a} - \Gamma^a_{ba,c} +
                \Gamma^r_{bc}  \Gamma^a_{ra} -
                \Gamma^a_{br}  \Gamma^r_{ca}  \nonumber  \\
 &=&
 \thbyth{-2(J_1^2+J_{11} )}{-J_{12} }{-J_{13} }{}{0}{-e^{2J}( J_{11}
   + 2 J_1^2) - 2 J_{23} }{}{}{0} , \end{eqnarray}
where the subscripts following commas denote derivatives with respect
to~$g,F$ and~$\bar{F}$ respectively.
The Ricci tensor~$R$ is symmetric, thus we have written down
only the upper triangle in 3.9.
Fortunately, there are symbolic manipulation programs
available that allow one to calculate the Ricci tensor
from the metric tensor in a matter of moments.

The curvature tensor vanishes iff the~$g,F, \; \bar{F}$ system
is a diffeomorphism of the Cartesian system. In 3
dimensions this condition is satisfied iff the Ricci tensor
vanishes.\fot{Wei} Thus
since~$J_{12} = J_{13} = 0$, we need:
\begin{equation} J = \ln G(g) + \ln M(F,\bar{F}) . \end{equation}
Since~$J_1^2 + J_{11} = 0$,
then up to scaling and translation we have
\begin{equation} \mbox{
\begin{tabular}{lll}
 \hspace{20ex}  &\mbox{\underline{A.} } $G=1/\sqrt{2}$ &or \hspace{30ex} \\
  &\mbox{\underline{B.} } $G= g/\sqrt{2}$. & \\
\end{tabular} }
\end{equation}
Conditions A and B ensure that all the components of the Ricci
tensor (3.11) vanish except for the (2,3) component.

\underline{The case A}
leads to~$(\ln M)_{23} = 0$ (since~$J_1=0$ in 3.11)
implying~$2J = - \ln 2 + \ln W'(F) +
\ln \bar{W}' (\bar{F})$, for some~$W, \; \bar{W}$ which
are arbitrary functions of~$F,\bar{F}$ respectively
(the prime~$'$ denotes derivative).
Then~$ 2e^{2J}
= W'(F)  \bar{W} '(\bar{F}).$ Thus the line element may be written
as~$ {\rm d}s^2 =  {\rm d}g^2 + {\rm d}W  {\rm d} \bar{W} $.
Thus we may define~$ x $ and~$y$ by~$ W =x +iy$ so that we
identify~$g$ with the Cartesian variable~$z$.
That is, we may
pick axes so that~$g=z$ and~$F$ is any complex analytic function
of~$W=x+iy$. Thus \underline{for case A},~$
{\rm d}s^2 =  {\rm d}z^2 + {\rm d}x^2 +  {\rm d}y^2 $, and
\begin{equation} \vec{c} =  e^{iz} \vec{\nabla} \left( F(x+iy) \right)  =
                e^{iz} F' \cdot (1,i,0) . \end{equation}
For the simple helical field mentioned in section 2,
we would have~$ F(x+iy) = -i (x+iy)$
and~$ \vec{c} =  e^{iz}  (-i,1,0) . $
Curiously it seems to have been Beltrami\fot{Bel}
who first wrote down that the Trkal equation
was solved by the real part of 3.12; unfortunately
this line of thinking has never been developed
thoroughly until this article.

\underline{The case B}
We wish the 23 component of the Ricci tensor to vanish:
\begin{equation} e^{2J}( J_{11}
   + 2 J_1^2) + 2 J_{23} =0 . \end{equation}
We substitute~$J  = - \frac{1}{2} \ln 2 + \ln g + \ln M(F,\bar{F})$
(this ensures that all the other components of
the Ricci tensor vanish) into 3.7 and 3.15 giving
\begin{equation} {\rm d}s^2 =  {\rm d}g^2 +
      g^2 \left( M(F,\bar{F}) \right)^2
                {\rm d}F  {\rm d}\bar{F} , \end{equation}
with
\begin{equation} - M^2 = 4 \pddr{\ln M}{F}{\bar{F}} . \end{equation}
As one might expect from the line element 3.14,
we can relate the case B to the
study of the spherical coordinates.

Equ. 3.15 has fascinating properties.$^{\ref{Gel} - \ref{JosLun} }$ It
arises in the study of heat propagation,
movement of plasmas and traces its origins back to
Emden, Darboux,\fot{Dar} and Liouville\fot{Lio} as well.
In different contexts, it bears the names as Emden-Fowler, Arrhenius
and Thomas-Fermi equations.
First we note that it
is equivalent to the two-dimensional
equation~$ \triangle \Psi = - e^{\psi}$
(using~$M= \frac{1}{\sqrt{2}} e^{\psi/2}$).
Second we note that if~$M(F, \bar{F})$ is a solution
to 3.15 then~$  | A'(F)| M(A, \bar{A})$ is also a solution
to 3.15 for any arbitrary complex analytic function,~$A(F)$
(or~$\psi (F, \bar{F})$ is a solution implies that
 $\rightarrow \psi ( A(F), \bar{A} ( \bar{F} ))
    + \ln (A(F)) + \ln (\bar{A} ( \bar{F} ))$ also is
a solution).

Thus introduce~$A, \bar{A}$ so that
\begin{equation} M^2 =
 A'(F) \bar{A}'(\bar{F})/ \cosh ^2( \frac{ A + \bar{A}}{2} ). \end{equation}
This solves 3.15 for arbitrary complex analytic functions~$A, \; \bar{A}$.
That any solution to 3.15 may be expressed in the form
of 3.16 was originally given by Liouville\fot{Lio}
(who expressed his solution with respect to a variable corresponding
to the exponential of the function~$A$.)
For example, setting~$A = b( \ln c +   \ln F ) $,
where~$b$, and~$c$ are real constants
yields~$M =   2 b / ( R ( (c R)^b + (cR) ^{-b} ) )$,
where~$R^2 = F \bar{F}$.
Solutions of this type correspond to the much discussed radial solutions
in flame propagation.\fot{Gel}
On the other hand setting~$A = -i b( \ln c +   \ln F ) $,
yields~$M = 1/(R \mbox{ cosh }( \mbox{ arg } F)), $~where
arg takes the argument of a complex number.

Next define~$ \cos \theta = \tanh \frac{ A + \bar{A}}{2}$
and~$ \phi = i (A- \bar{A})/2$.
Then it is not difficult
to show that~$M^2  {\rm d}F  {\rm d}\bar{F} =
{\rm d}\theta ^2 + \sin^2 \theta {\rm d}\phi ^2.$
Thus
\begin{equation} {\rm d}s^2 =  {\rm d}r^2 +
      r^2 ({\rm d}\theta ^2 + \sin^2
                \theta {\rm d}\phi ^2),\end{equation}
and we recover the spherical coordinate system.
Since~$A$ was a completely arbitrary function of~$F$,
then conversely~$F$ is a completely arbitrary
function of~$A$. So
due to the definitions of~$\theta,$~$ \phi$ we have
\begin{equation} F_{\rm spherical} =
    F ( - \mbox{arc tanh}(\cos \theta) + i \phi ). \end{equation}
We should require also that either~$F$ be a~$2 \pi$-periodic
function of~$\phi$, or a (complex) scalar multiple of the identity function.
This ensures that theresulting vector fields are single valued.
Notice that~$F$ is anihilated by the
operator~$ \partial _\theta + i \frac{1}{\sin \theta} \partial _\phi$,
which may be used to factor the angular momentum operator.
And so
\begin{equation} \vc = \frac{ \; e^{ir}}{r \sin \theta}
    \left[ \hat{\theta} + i \hat{\phi}
                    \right]  F'  .\end{equation}
A typical case is given in Figures 4 and 5.
Choosing~$F(a) = \cos(-ia)$ and using 3.18 and 3.19
yields~$\vv = \frac{-1}{r \sin^2 \theta} \left\{
    \hat{\theta} [ \cos r \cos \phi \cos \theta + \sin r
                    \sin \phi ] +
    \hat{\phi} [ \cos r \sin \phi - \sin r \cos \phi
                    \cos \theta ] \right\} $.
Thus even a very simple function chosen for~$F$ may yield
a rather complicated expression for~$\vv$.

In summary, there are precisely two cases:
A. the Cartesian case (3.12), where the screw direction is in
the direction of a Cartesian variable and~$F$ is an analytic
function of the other two Cartesian variables, and
B. the spherical case (3.19), where the screw direction is in
the radial direction of a spherical coordinate system
and~$F$ is an analytic function of the
variable~$ (- \mbox{arc tanh}(\cos \theta) + i \phi )$,
with the additional requirement that the resulting
vector field be single valued.

It is straightforward to calculate the helicity
for either cases A or B.
We find
\begin{eqnarray}  \mbox{Case A:~~~}  h &=&  v^2 =
    \frac{1}{2} \vec{\nabla} F \cdot \vec{\nabla} \bar{F} =
              F' \bar{F}' , \nonumber \\
       \mbox{Case B:~~~}  h &=&  v^2
    =  \frac{1}{2} \vec{\nabla} F \cdot \vec{\nabla} \bar{F}
     = F' \bar{F}'    /(r^2 \sin ^2 \theta ) , \end{eqnarray}
where~$F'$ depends on either A.~$x+iy$ or
B.~$- \mbox{arc tanh}(\cos \theta) + i \phi $.
In both cases~$\triangle F = 0$,
the 3-dimensional Laplacian of~$F$ is zero for either class.
In case A, actually the two-dimensional Laplacian is of course zero, and
in case B, the ``angular momentum part'' of the Laplace operator
 acting on~$F$ is zero.

As a final remark,
notice from 3.20 that there are no oscillations in the helicity
(equivalently velocity magnitude)
in the direction of increasing~$g$, (that
is~$\hat{z}$ or~$\hat{r}$.)
This is related to the fact that~$\vec{c}$ is
a complex null vector~$\vec{c} \cdot \vec{c} = 0 $ ,
so that only the cross terms~$ \vec{v} ^2 = 0 $
survive in calculating~$v^2$.

{}From 3.20, we notice that nothing precludes that the helicity
should vanish at exceptional points. Note also
that the integrated helicity over any infinite
domain of~$R^3$ (for the complex Trkalian flows)
will always be infinite. In that sense,
these solutions lack a certain
natural global character.

%%%%%%%%      SECTION 4 %%%%%%%%%%%%%%%%%%%%%%%%%%%%%%%%%%%%%%
%%%%%%%%                %%%%%%%%%%%%%%%%%%%%%%%%%%%%%%%%%%%%%%
%%%%%%%%                %%%%%%%%%%%%%%%%%%%%%%%%%%%%%%%%%%%%%%
\section{Boundary Conditions }
\renewcommand{\theequation}%
{\arabic{section}.\arabic{equation}}
\setcounter{equation}{0}
\hspace{4ex}
Bj\mbox{\o}rgum has shown that in regions free
from singularities, that if
either~$\vv \cdot {\bf \vec {n}}=0$
or~$\vv \times {\bf \vec {n}}=0$,
where~$\vv$ is a solenoidal Beltrami field
and~${\bf \vec {n}}$ is the normal to some bounding surface,
that it necessarily follows that
the vector field~$\vv$ must vanish identically.
We are then able to discuss which
types of boundary conditions the
{\bf Complex Trkalian Fields} may obey.

Since the type A complex
Trkalian fields (with screw direction along a Cartesian axis),
are solenoidal, they cannot satisfy the aforementionned
natural boundary conditions.
The type B Complex Trkalian fields always have
a singularity interior to any region
containing the origin so that the rigorous result
of Bj\mbox{\o}rgum does not apply.
Moreover these fields always have a
vanishing component in the radial direction, so that
one may pick a boundary (~$r=$~a constant) that encloses
a simply connected region in such a way that the
vector field lies tangent to such a surface.
However, there is a line singularity passing through the
center of the sphere, leading to an
infinite value of the integrated helicity.
In conclusion then, we see that no finite amplitude complex
Trkalian field in a simply connected domain
may satisfy the physically interesting
boundary conditions~$\vv \cdot {\bf \vec{n}} =0$ on some surface.
In multiply connected regions, Trkalian fields satisfying physically
interesting boundary conditions may exist.
%although not as much is known rigorously.

%%%%%%%%      SECTION 5 %%%%%%%%%%%%%%%%%%%%%%%%%%%%%%%%%%%%%%
%%%%%%%%                %%%%%%%%%%%%%%%%%%%%%%%%%%%%%%%%%%%%%%
%%%%%%%%                %%%%%%%%%%%%%%%%%%%%%%%%%%%%%%%%%%%%%%
\section{Bj\mbox{\o}rgum and Godal's
           Work on Beltrami Fields }
\renewcommand{\theequation}{\arabic{section}.\arabic{equation}}
\setcounter{equation}{0}
\hspace{4ex}
In this section, we discuss Bj\mbox{\o}rgum et. al.'s work
(1948, 1951, 1952) and intepret his results in the same framework
as ours.
In his first long paper on Beltrami fields,\fot{Bjo} Bj\mbox{\o}rgum points
writes that the Beltrami vector fields are precisely those
vector fields which may be written as:
\begin{equation} \vv = \frac{ \; \vec{\nabla} q_1 \times \vec{\nabla} q_2 }{
\Omega }
      = q_1 \vec{\nabla} q_2 + \vec{\nabla} q_3 \end{equation}
where~$\Omega$ is the torsion function. This leads to
\begin{equation}  \pdr{\vec{r}}{q_a} \cdot \pdr{\vec{r}}{q_b}
        = \thbyth{R}{S}{0}{S}{T}{q_1 \frac{\Omega}{H}}{0}%
{q_1 \frac{\Omega}{H} }{ \frac{\Omega}{H} }_{ab}, \end{equation}
where~$R,S,T,M$ are themselves functions,
and~$T$ may be found from:
\begin{equation} T = S^2/R + \frac{q_1^2 \Omega}{H} + \frac{1}{\Omega  H R}.
\end{equation}
(The determinant of the matrix 5.2 is~$1/H^2$,
where~$H= \vec{\nabla} q_1 \cdot \vec{\nabla} q_2 \times \vec{\nabla} q_3$
is the Jacobian of the transformation from the~$q$'s to
the Cartesian variables.)
One may demonstrate Eq. 5.2, using the identity:
\begin{equation} \pdr{\vec{r}}{q_3} =
  \frac{\vec{\nabla} q_1 \times \vec{\nabla} q_2}{\vec{\nabla} q_1 \cdot
\vec{\nabla} q_2 \times \vec{\nabla} q_3},
      \end{equation}
which holds for any three functionally unrelated functions~$q_1,
\; q_2, \; q_3 $.
Using 5.4, the Beltrami equation (Eq. 5.1)
may be written
\begin{equation} \pdr{\vec{r}}{q_3} =
    \Omega \left( q_1  \pdr{\vec{r}}{q_3} \times
           \pdr{\vec{r}}{q_1}  + \pdr{\vec{r}}{q_1}
        \times \pdr{\vec{r}}{q_2} \right) . \end{equation}
Equation 5.2 then follows directly.
Bj\mbox{\o}rgum, however does not ensure that the above
metric tensor given by 5.2 yields a zero curvature tensor.
This {\bf\em must} be the case if the~$q_1, q_2$ and~$q_3$ coordinates
are to be themselves functions of Cartesian coordinates,
as we argue in our paper.

If one uses the representation given in Eq.~A.7:
\begin{equation}  \vv = \frac{1}{2} (e^{iy_1} \vec{\nabla} y_2
      + e^{-iy_1} \vec{\nabla} y_3) , \end{equation}
where~$y_1 = q_2$ is real and~$y_2$ and~$y_3$ are complex conjugates,
then Bj\mbox{\o}rgum's statement is equivalent to the following.
The Beltrami fields are precisely those vector
fields where:
\begin{equation}  \pdr{\vec{r}}{y_a} \cdot \pdr{\vec{r}}{y_b}
        = \thbyth{B}{-\frac{i}{2}e^{iy_1}A}{\frac{i}{2}e^{-iy_1}A}%
{-\frac{i}{2}e^{iy_1}A}{e^{2iy_1}D }{P}%
{\frac{i}{2}e^{-iy_1}A}{P}{e^{-2iy_1}D}_{ab}, \end{equation}
where
\begin{eqnarray}  B &=&  \frac{1+ A^2 \Omega H}{ H \Omega R}, \nonumber \\
       D &=&  \frac{\Omega - RH}{4 H}, \nonumber \\
       P &=&  \frac{\Omega + RH}{4 H}, \end{eqnarray}
are real functions. Eq. 5.7 may
be used as a starting point for the Ricci tensor
calculation. Given~$\Omega$ and~$ H$, there are two functions
(say~$R$ and $S$) that characterize the matrix (5.2).
Similarily, two functions characterize the matrix given
by 5.7 (say~$R$ and $A$). The function~$A$ is
related to~$S$ by
\begin{equation}   A = S +R q_3. \end{equation}
The determinant of the matrix 5.7 is~$ [D+P]
[B(D-P) + a^2/2] = \frac{-1}{4 H^2} = \frac{1}{H'^2} .  $
This is because~$H'= \vec{\nabla} y_1 \cdot \vec{\nabla} y_2 \times
\vec{\nabla} y_3 =
-2i \vec{\nabla} q_1 \cdot \vec{\nabla} q_2 \times \vec{\nabla} q_3 = -2i H$ as
may be found from calculating~$\vv \cdot \vec{\nabla} \times \vv$ from
5.1 and 5.6.
To proceed from 5.2 to 5.7, one needs
\begin{equation}  \pdr{\vec{r}}{y_a} \cdot \pdr{\vec{r}}{y_b}  =
      \pdr{q_i}{y_a}
          \pdr{\vec{r}}{q_i} \cdot \pdr{\vec{r}}{q_j}
            \pdr{q_j}{y_b} =
      Z_{ia}
          \pdr{\vec{r}}{q_i} \cdot \pdr{\vec{r}}{q_j}
                  Z_{jb},
      \end{equation}
with
\begin{equation}
      Z_{ia}  = \pdr{q_i}{y_a} =
       \thbyth{q_3}{1}{-q_1}%
{\frac{-i}{2} e^{iy_1}}{0}{\frac{1}{2}e^{iy_1}}%
{\frac{i}{2} e^{-iy_1}}{0}{\frac{1}{2}e^{-iy_1}}. \end{equation}
The above matrix~$[Z]$ may be found straightforwardly by
using~$y_1 = q_2, \; y_2 = (q_3 + iq_1)e^{-i q_2}, \;
          y_3 = ( q_3 - iq_1)e^{+i q_2}$.
The Beltrami equation in terms of the~$y$ variables
then may be written
\begin{equation} e^{iy_1} i \pdr{\vec{r}}{y_3}
          + i e^{-iy_1} \pdr{\vec{r}}{y_2}
    =  \Omega \left( e^{iy_1} \pdr{\vec{r}}{y_3}
             \times \pdr{\vec{r}}{y_1}
        + e^{-iy_1}\pdr{\vec{r}}{y_1}
              \times \pdr{\vec{r}}{y_2}  \right). \end{equation}

For complex Trkalian fields it is straightforward to show
in that case that one must have~$A=D=0$ and~$B=1$ (which
specifies the torsion function).
This matches the coefficients of~$e^{iy_1}$ and~$e^{-iy_1}$
above in Eq. 5.12, or equivalently,
discards the oscillating terms of the matrix, 5.7.
At that point one is left with the matrix described below
Eq. 3.6:
\begin{equation}  \pdr{\vec{r}}{y_a} \cdot \pdr{\vec{r}}{y_b}
        = \thbyth{1}{0}{0}{0}{0}{\frac{1}{iH'}}%
{0}{\frac{1}{iH'}}{0}_{ab}, \end{equation}
The curvature condition then ensures a complete solution.
In future work we will proceed to examine less restrictive
cases than the complex Trkalian case and calculate curvature
more directly from Eq 5.7.
One might suspect that
the complex Trkalian fields hold a very prominent place in
the theory of the Beltrami fields, since they seem to take the
elements of the matrix 5.7, which are most robust
in some sense. As we show in later work the
complex Trkalian fields may be shown to yield precisely
the linearly polarized TEM solutions\fot{PRBTEM} of
electromagnetic wave theory.

%%%%%%%%      SECTION 6 %%%%%%%%%%%%%%%%%%%%%%%%%%%%%%%%%%%%%%
%%%%%%%%                %%%%%%%%%%%%%%%%%%%%%%%%%%%%%%%%%%%%%%
%%%%%%%%                %%%%%%%%%%%%%%%%%%%%%%%%%%%%%%%%%%%%%%
\section{Conclusions - Discussion}
\renewcommand{\theequation}{\arabic{section}.\arabic{equation}}
\setcounter{equation}{0}
\hspace{4ex}
In this paper we have introduced a new method for
constructing true three dimensional vector field solutions
to vector pde's of mathematical physics.
As opposed to axisymmetric flows,
which are the dominant solutions to fluid mechanical
equations listed in texts, true three-dimensional
flows should have a finite value of helicity~$\vv \cdot
\vec{\nabla} \times \vv$.
It is clear that the Clebsch representation
is then the most convenient:~$\vv = f \vec{\nabla} g + \vec{\nabla} h$.
That is because non-zero helicity fields may always
be expressed in such a manner\fot{Fla}
and can never be expressed with fewer than three
Clebsch functions.
Given the representation, however,
the question remains how to construct these functions (f,g,h).

Our approach
then is to represent the appropriate
vector field with the appropriate Clebsch functions
and then to construct the relations between
the elements of a metric tensor,~$q_{ij}$, using the
original pde. After
having constructed a metric tensor consistent with
the equations one wishes to solve,
one computes the Ricci tensor.
If the Clebsch functions are to be functions of the
Cartesian variables, then (in three dimensions)
the Ricci tensor must necessarily vanish.
After solving the partial differential equations for the vanishing
of the Ricci tensor, one finally searches for the class of
transformations of the metric tensor back to Cartesian coordinates.
By this means, the integrity of the good coordinates -
the Clebsch functions - remains intact.

In particular in this manuscript, we have undertaken the first
steps in a systematic
classification of Beltrami\fow{MclPir}{ConMaj} and Trkalian fields
which are thought to be important constituents of
high Reynolds number flow fields.
For example the so-called ABC (Trkalian) flow has
been extensively studied\fot{Bal3} particularily for its
chaotic properties.\fow{LibSiv}{NakHatKam}
Note that, although
it is always possible to write down a formal solution
to the Trkalian equation as in equ. 2.5 in terms of Fourier
modes, the use of Clebsch functions, which we advocate here,
has many advantages.
For example, when one Fourier decomposes a problem one implicitly
assumes that the Cartesian coordinates are a natural coordinate
system for the problem.
As we have discussed above this is not necessarily the case.

In this particular article the analysis
proceeds by considering the Beltrami fields
as real parts of complex vector fields, and then
considering complex vector fields of a certain type.
Complex vector fields play a prominent role in many
areas of mathematical physics including
electromagnetism (see Bateman\fot{Bat})
and minimal surfaces (see Struik\fot{Str} or Osserman\fot{Oss})
 In this paper, we examine complex vector fields,
which satisfy the Trkalian equation (1.3), and may be written
in the form~$e^{ig} \vec{\nabla} F$, with~$g$ real and~$F$ complex
(that is, we restrict the problem enough
in order to give a complete solution).
Both the real and imaginary parts of such a vector field
will then be Trkalian.
The two classes of solutions
to the complex Trkalian equation with the complex vector field of the
above form are given in section 3.
The class of solutions where~$\vv$ is a 2-d potential
field with a screw direction in a Cartesian
direction has been studied before (Bj\mbox{\o}rgum's uniplanar Trkalian
flows).
The class of solutions where~$\vv$ is a ``zero angular momentum
field'' with a screw direction in the spherical radial
direction does not appear in the Beltrami literature.
Examples of both types of fields
are presented in the figures.

Although the Clebsch representation of the vector field
has found fruitful application in
fluid mechanics\fow{Cle}{Tru}, thermodynamics and
more generally in variational
principles,\fow{MatVed}{HolKim}
the authors still maintain that this representation has
been underexploited.
We demonstrate in this article that the method we have
used employing Clebsch functions has much practical
merit in producing new solutions to
partial differential equations.
The same method will be used in further articles
on fluid mechanics, and wave equations.

The author would like to express his deep gratitude to
R. M. Kiehn at the University of Houston for teaching him
about differential forms, the intrinsic merit of Clebsch
functions, and a topological approach to Physics
and for sharing his results prior to publication.\fot{Kie}
The author also thanks Professor Giles Auchmuty of the University
of Houston for discussions and for pointing out several
references.

%%%%%%%%   APPENDIX A   %%%%%%%%%%%%%%%%%%%%%%%%%%%%%%%%%%%%%%
%%%%%%%%                %%%%%%%%%%%%%%%%%%%%%%%%%%%%%%%%%%%%%%
%%%%%%%%                %%%%%%%%%%%%%%%%%%%%%%%%%%%%%%%%%%%%%%
\setcounter{equation}{0}
\renewcommand{\theequation}{ A.\arabic{equation} }
\newpage
\appendix
\section{ Vector Fields and Their Topological Duals}
\hspace{4ex}
All the following arguments are local for a region where
the helicity is of constant sign (non-zero).
\begin{Proposition}
Every vector field,~$ \vec{v}$, may be written as the real part
of a complex vector field of the following form:
\begin{equation} \vec{ c} =   e^{ig} \vec{\nabla} F , \end{equation}
where~$g$ is a function with domain~${\cal R}^3$
and range~${\cal R}$, and~$ F$ has domain~${\cal R}^3$
and range the complex numbers.
\end{Proposition}
{\bf Proof:} One knows that it is always possible to construct the
Monge potentials~$f,g,h$ so that a vector field~$\vv$ may be expressed in
the decomposition due to Clebsch:\fot{Fla}
\begin{equation} \vv = f \vec{\nabla} g + \vec{\nabla} h. \end{equation}
Now define
\begin{eqnarray}  F   &\equiv& (h + i f) e^{-ig}, \\
  \bar{F}  &\equiv& (h - i f) e^{+ig}. \end{eqnarray}
Or equivalently,
\begin{eqnarray}  f &=& \frac{-i}{2} (F e^{ig} - \bar{F} e^{-ig}), \\
       h &=&  \frac{1}{2} (F e^{ig}  + \bar{F} e^{-ig}).   \end{eqnarray}
Now we may write~$\vv$ as
\begin{equation} \vv =  \frac{1}{2} (e^{ig} \vec{\nabla} F + e^{-ig}
\vec{\nabla} \bar{F})
           , \end{equation}
from which 1 follows.

\begin{Proposition}
The real part of~$ c$,~$\vv$, and
the imaginary part of~$ c$,~$\vec{w}$ satisfy:
\begin{eqnarray}  \vec{\nabla} \times \vv + \vec{\nabla} g \times \vec{w} &=& 0
 , \\
       \vec{\nabla} \times \vec{w} - \vec{\nabla} g \times \vv &=& 0  , \\
                \vv \times \vec{w}  \ne 0 \end{eqnarray}
We say that any two vector fields that
satisfy A.8 - A.10 for some
function,~$g$, are said to be~{\bf dual} to one another with respect to~$g$.
\end{Proposition}
Equs. A.8 - A.10 are easy to show beginning from
\begin{equation} \vec{w} =  -h \vec{\nabla} g + \vec{\nabla} f. \end{equation}
In particular it may be shown from the above that dual vector fields
satisfy
\begin{eqnarray}  \vec{w} \cdot ( \vec{\nabla} \times \vv) &=& 0  , \\
       \vv \cdot ( \vec{\nabla} \times \vec{w}) &=& 0  , \\
       \vv \cdot ( \vec{\nabla} \times \vv) &=& \vec{w} \cdot ( \vec{\nabla}
\times \vec{w}) \end{eqnarray}
\begin{equation}      \vec{\nabla} \times \left(
  \frac{ (\vv \times \vec{w} ) \cdot  (\vec{\nabla} \times \vv) }{
          (\vv \times \vec{w} ) \cdot (\vv \times \vec{w} ) } \vv
  -  \frac{ (\vec{w} \times \vv ) \cdot  (\vec{\nabla} \times \vec{w}) }{
          (\vv \times \vec{w} ) \cdot (\vv \times \vec{w} ) } \vec{w} \right.
\nonumber \\
  \left.  - \frac{ \vv \cdot  (\vec{\nabla} \times \vv) }{
  (\vv \times \vec{w} ) \cdot (\vv \times \vec{w} ) } \vv \times \vec{w}
              \right)   = 0 . \end{equation}
The terms in the parenthesis of A.15 constitute a
solution for~$-\vec{\nabla} g$ from A.8 and A.9.

A complete classification of vector fields in this fashion
does not seem to be known.
For example, one can also take a vector field in~${\cal R}^4$
and write it as the real part of a zero helicity
vector field with domain the complex numbers. One can also
write a vector field in~${\cal R}^8$
as the real part of a zero helicity
vector field with domain the quaternions.
For the dimensions 5, 6 and 7 however no obvious standard form
exists, although clearly these vector fields may be thought
of as the real part of zero helicity quaternionic fields of
a restricted type (the same way that any real vector field
may be thought of as the real part of a complex vector field
satisfying,~$\vec{\nabla} \times \vc = i \vec{\nabla} g \times \vc$.)

Dual concepts are prevalent in mathematical litaerature.
Typically one hopes to represent a function (or vector of functions)
as the real part of a complex function with slightly
nicer properties.
The complex function uses the same functions as the real part
as building blocks. Thus one writes a harmonic function
as the real part of an analytic function (which depends on
a single variable, but where this variable belongs to the complex
field).
Similarily one writes the panharmonic function as the real part of
a~$\mu$ regular function (see Duffin\fot{Duf}).

%%%%%%%%   APPENDIX B  %%%%%%%%%%%%%%%%%%%%%%%%%%%%%%%%%%%%
%%%%%%%%               %%%%%%%%%%%%%%%%%%%%%%%%%%%%%%%%%%%%%%
%%%%%%%%               %%%%%%%%%%%%%%%%%%%%%%%%%%%%%%%%%%%%%%
\section{ The Proposed Method Employed in a Familiar Problem }
\setcounter{equation}{0}
\renewcommand{\theequation}{ B.\arabic{equation} }

\hspace{4ex} We give a very simple example of the general procedure we
are outlining. Consider the case of incompressible, time
independent, irrotational
flow past an obstacle in two dimensions.
We wish to study a potential function,~$\phi$,
a stream function~$\psi$, and a complex
potential~$W = \phi + i \psi$.
Let's try to write a differential element of length in
terms of the functions~$\phi$ and~$\psi$.
Since~$\phi$ and~$\psi$ are respectively the real
and imaginary part of the complex potential~$W$,
they satisfy the Cauchy-Riemann
relations:~$ \pdr{\phi}{x} = - \pdr{\psi}{y} ; \;
\pdr{\phi}{y} = \pdr{\psi}{x} $,
or more compactly
\begin{equation}  \vec{\nabla} \phi = \hat{z} \times \vec{\nabla} \psi .
\end{equation}

Now one may straightforwardly
calculate
\begin{eqnarray}   \vec{\nabla} \phi \cdot \vec{\nabla} \phi &=&
          \vec{\nabla} \psi \cdot \vec{\nabla} \psi, \\
        \vec{\nabla} \phi \cdot \vec{\nabla} \psi &=& 0 .\end{eqnarray}
For shorthand, let~$y^1 = \phi, \; y^2 = \psi $.
Then~$g^{lk} = \pdr{y^l}{x^p} \pdr{y^k}{x^p} =
\twbytw{1/a^2}{0}{0}{1/a^2} $.
The matrix
\begin{equation} g_{ij} =  \pdr{x^p}{y^i} \pdr{x^p}{y^j} =
\twbytw{a^2}{0}{0}{a^2} \end{equation}
is the matrix inverse of~$g^{lk}$.

1. The  first step of the calculational procedure is then
complete: for the equations of irrotational, incompressible
flow (Cauchy-Riemann conditions), \underline{the appropriate
non-vanishing} \underline{components of the metric tensor}
are identified.
We are then left with a line element (or metric tensor)
given by
\begin{equation} {\rm d}s^2 =  a^2 \left( {\rm d}\phi ^2 +  {\rm d}\psi ^2
  \right) , \end{equation}
where~$a =a ( \phi, \psi ) $.
Of course,~B.1 immediately implies B.5 due to
conformality of a complex anayltic transformation;
one needs virtually no calculation.
In the approach taken here, {\bf flow equations like B.1
are interpreted as geometric constraints
(on the metric function).}
Note that in his papers Bj\mbox{\o}rgum calculated correctly
the non-vanishing components of the metric tensor.
He never ensured however that the curvature was zero.
Applying this spirit to the current problem, Bj\mbox{\o}rgum
would have halted at this point.

2. Now we calculate \underline{the Ricci tensor}
given the metric function B.4:
\begin{equation} g_{ij}  = \twbytw{a^2}{0}{0}{a^2}_{ij} .\end{equation}
Let~$A = \ln a$. Then
\begin{eqnarray}  \Gamma^i_{kl}
            & \equiv & \frac{1}{2} g^{im}
   \left( \pdr{g_{mk}}{y^l} +  \pdr{g_{ml}}{y^k}
          -  \pdr{g_{kl}}{y^m} \right)  \\
    &=& A,_l \delta_{ik} +
     A,_k \delta_{il} - A,_i \delta_{lk} \;\;\; , \end{eqnarray}
where the indices following commas denote derivatives.
Now the Ricci tensor is given by
\begin{equation} R_{bc} \equiv \Gamma^a_{bc,a} - \Gamma^a_{ba,c} +
                \Gamma^r_{bc}  \Gamma^a_{ra} -
                \Gamma^a_{br}  \Gamma^r_{ca} =
        -( A,_{11} + A,_{22} )
              \twbytw{1}{0}{0}{1}_{bc} .\end{equation}
Now we may express the functions~$ \phi $
and  $\psi $ in terms of the Cartesian
variables if and only if the Ricci tensor vanishes identically.
Thus we need the function~$ \ln a $ to be harmonic
with respect to the functions~$ \phi$ and ~$\psi$.
Actually this is quite obvious since
\begin{equation} {\rm d}s^2 =   {\rm d}z   {\rm d} \bar{z} =
   \der{z}{W} \: \der{\bar{z} }{\bar{W} }{\rm d}W   {\rm d}
        \bar{W} ,\end{equation}
where~$z=x+iy$, so that~$\pddr{\mbox{ } }{ W}{\bar{W}}  \ln
   \left( \der{z}{W} \der{\bar{z} }{\bar{W}} \right)  = 0$.
The above ``$A$'' corresponds to~$\ln \der{z}{W}$.

3. \underline{Boundary Conditions}  \\
The actual choice for~$\ln a$ must be determined by
the boundary conditions of an actual two
dimensional flow problem.
For flow past an obstacle,
we wish to map (part of)
the real axis of the variable~$W = \phi + i \psi$ to
the boundary  of the obstacle.
The upper half plane is then mapped to the exterior of the
obstacle in the upper half plane.
Once~$a( \phi, \psi) $ is determined, then we may try to
calculate the Cartesian variables by means of
the eikonal like equations
\begin{equation}  (\pdr{x}{\phi}) ^2 + (\pdr{x}{\psi}) ^2=
      (\pdr{y}{\phi}) ^2 + (\pdr{y}{\psi}) ^2=
                      a^2(\phi,\psi)  . \end{equation}

There is no new result outlined in the approach above,
we merely demonstrate the method, which we use to construct
Clebsch functions in this paper, to a more familiar problem.

4. \underline{Examples} \\
Symmetrical flow past a unit disc with unit speed far downstream.
We need a function mapping (part of) the real line
to the boundary of a unit disc. This is well
known:~$z \equiv x+ iy = e^{i \arccos \frac{W}{2} }
= (W  + \sqrt{W^2 - 4})/2 $
We then calculate~$a^2 = \left| \der{z}{W} \right| ^2 =
\frac{1}{ 4 }
\left| 1 + \frac{W}{ \sqrt{ W^2 - 4 } } \right| ^2 $.
So~${\rm d}s^2 = a^2 ( {\rm d}\phi ^2 + {\rm d}\psi ^2 )$,
where~$\phi,\psi$ are the real and imaginary parts of~$W$.

For symmetrical flow past a plate of width 4
aligned perpendicular to the flow along the~$x$ direction,
we have~$z = \sqrt{W^2 -4},$ and
we calculate~$a^2 = \left| \der{z}{W} \right| ^2 =
\frac{\phi ^2 + \psi^2}{ \sqrt{ (\phi^2 + \psi^2 - 4)^2
     + 16 \psi^2 } } $.
For both plate and disc, there is a singularity in~$a$
when~$\phi =2 ; \psi =0$.

Conversely given the function~$a^2$, (where~$\ln A$ is harmonic
in~$\psi, \phi$) then one may use B.11 to reconstruct the
boundary.

Also note that since the streamlines correspond to contours of
constant~$\psi$, then roughly the function~$\psi$ plays
the role of an energy for a 1 degree of freedom Hamiltonian system,
whereas the function~$\phi$ plays the role
of a phase (or time) variable.

As is well known,
the harmonic functions are precisely those
functions which serve as nice coordinates of
orthogonal coordinate systems in 2-dimensions.
They are the ``isothermal parameters''\fot{Str} of the plane.
In a similar sense, the (Clebsch) functions we solve for in
this paper are  also the good coordinates for the problem at
hand: the building blocks of the complex Trkalian flows.

%%%%%%%%   REFERENCES   %%%%%%%%%%%%%%%%%%%%%%%%%%%%%%%%%%%%%%
%%%%%%%%                %%%%%%%%%%%%%%%%%%%%%%%%%%%%%%%%%%%%%%
%%%%%%%%                %%%%%%%%%%%%%%%%%%%%%%%%%%%%%%%%%%%%%%
\vspace{0.7in}
{\huge\bf References}

\vspace{5ex}
\newcounter{ref}
\begin{list}%
{[\arabic{ref}]}{\usecounter{ref}}

%**************************************************************************
\begin{center}    {\bf EXACT SOLUTIONS} \end{center}
%***************************************************************************
\item C. Y. Wang, \label{Wan1}
  {\em Annu. Rev. Fluid Mech.} {\bf 21} (1991) 159-177.\\
Exact Solutions of the Steady-State Navier-Stokes Equations

\item C. Y. Wang, \label{Wan2}
  {\em Acta Mechanica} {\bf 81} (1990) 69.
Exact Solutions of the Navier-Stokes Equations: the generalized
Beltrami flows, review and extension

\item C. Y. Wang, \label{Wan3}
  {\em Applied Mechanics Review} {\bf 42} (1989) S629.
Exact Solutions of the unsteady Navier-Stokes Equations

%***********************************************************************
\begin{center}    {\bf MHD } \end{center}
%***********************************************************************
\item D. Montgomery, J. P. Dahlberg, L. Phillips,
        and M. L. Theobald,  \label{MonDahPhiThe} \\
Minimum Dissipation States and Vortical Flow in MHD

\item G. Remorini, \label{Rem}
\underline{Rendiconti di Matematica, Serie VII},
Volume 9, Roma (1989), 45-52. \\
 Sui Moti di Beltrami in Magnetofluidodinamica
 Exist Beltrami flows in MHD

\item R. G. Storer, \label{Sto}
   {\em Plasma Physics} {\bf 25} (1983) 1279-1282.\\
Spectrum of an Exactly Soluble
Resistive Magnetohydrodynamic Model  \\

\item D. Dritschel, \label{Dri}
{\em J. Fluid Mech.}, {\bf 222} (1991) 525-541.  \\
Generalized Helical Beltrami Flows in Hydrodynamics
               and Magnetohydrodynamics

\item Viktor Trkal, \label{Trk}
  {\em Casopis pro p'estov'an'i matematiky a fysiky}
         {\bf 48} (1919)  302-311. \\
Pozn\'{a}mka k hydrodynamice vazk\'{y}ch tekutin

\item P. R. Baldwin, \label{PRB1}
      {\em  Constructing Clebsch Potential for Vector Fields}
     submitted to {\em Journal of Physics} {\bf A}.

\item P. F. Nemenyi and R. C. Prim, \label{NemPri}
  {\em Proceedings of the Seventh International Congress for
 Applied Mechanics} {\bf 7} (1948) 300.

\item A. W. Marris, \label{Mar2}
   {\em Archive for Rational Mechanics and Analysis}
   {\bf 70} (1979) 47.  \\
 The Impossiblity of Steady Screw Motions
   of a Navier Stokes Fluid

\item P. R. Baldwin, \label{PRBTEM}
{\em Phys. Lett.} {\bf A189} (1994) 161. \\
A Classification Result for Linearly Polarized
Principal Electromagnetic Waves.

\item L. D. Landau and E. M. Lifshitz, \label{LanLif2}
     \underline{Classical Theory of Fields}
            (Pergamon Press, 1968).

\item O. Bj\mbox{\o}rgum, \label{Bjo}
 {\em Universitetet I Bergen Arbok} {\bf 1} (1951) 1. \\
On Beltrami Vector Fields and Flows. Part I:
A Comparative Study of Some Basic Types of Vector Fields.\\
See also
 {\em International Congress for Applied Mechanics} {\bf 7} (1948) 341. \\
On Some Three-dimensional Solutions of the Non-linear
Hydrodynamical Equations.

\item O. Bj\mbox{\o}rgum and T. Godal, \label{BjoGod}
 {\em Universitetet I Bergen Arbok} {\bf 13} (1952) 1. \\
On Beltrami Vector Fields and Flows
Part II: The Case When~$\Omega$ is Constant in Space

\item P. R. Baldwin, \label{PRBTE}
{\em The Classification of Linearly Polarized
Transverse Electric Waves} submitted to
{\em Journal of Rational Mechanics and Analysis}.

\item L. Crocco, \label{Cro}
  {\em ZaMM} {\bf 17} (1937) 1-7.
  Eine Neue Stromfunktion der Bewegung der Gase
mit Rotation.

\item B. L. Hicks, P. E. Guenther and R. H. Wasserman,
      \label{HicGueWas}
  {\em  Quarterly of Applied Mathematics} {\bf 5} (1947) 357. \\
New Formulations of the Equations For Compressible Flow \\

\item P. R. Baldwin, \label{PRB2}
{\em Topologically Dual Vector Fields and Clebsch
  Functions in~${\cal R}^3$ and~${\cal R}^4$},
to be submitted to {\em Physics Letters A}.

\item S. Weinberg, \label{Wei}
  \underline{Gravitation and Cosmology} \\
    \underline{Principles and Applications of the General Theory
of Relativity}
    (J. Wiley and Sons, N.Y., 1972).

\item E. Beltrami, \label{Bel}
 {\em Reale Istituto Lombardo di Scienze e Lettere} {\bf 22} (1889) 112-131. \\
Considerazioni Idrodinamiche

%***********************************************************************
\begin{center}    {\bf EMDEN-FOWLER PDEs} \end{center}
%***********************************************************************
\item I. M. Gelfand, \label{Gel}
  {\em Amer. Math. Soc Transl. (2)} {\bf 29} (1963) 295. \\
Some Problems in the Theory of Quasilinear Equations

\item H. Fujita, \label{Fuj}
  {\em Bull. Amer. Math. Soc.} {\bf 75} (1963) 295. \\
On the nonlinear equations~$\triangle u + e^u$~and~%
$ \pdr{v}{t}=\triangle v + e^v$

\item D. D. Joseph and T. S. Lundgren, \label{JosLun}
  {\em Arch. Rational Mech. Anal.} {\bf 49} (1972) 241.\\
Quasilinear Dirichlet Problems Driven by Positive Sources

\item Gaston Darboux, \label{Dar}
  \underline{Le\c{c}ons sur les Syst\`{e}mes
   Orhogonaux et les Coordonn\'{e}es Curvilignes}\\
  \underline{Le\c{c}ons sur la Th\'{e}orie G\'{e}n\'{e}rale
   des Surfaces: Premi\'{e}re Partie  }\\
       (Gauthier-Villars, Paris, 1889). \\

\item Liouville's Journal {\bf vol. 18} (1853) 71. \label{Lio}
 see  H. Bateman, \\
   \underline{Partial Differential Equations
  of Mathematical Physics} (1959) 166-169.

\item H. Flanders, \label{Fla}
   \underline{Differential Forms with Applications
        to the Physical Sciences} (Dover, N. Y.,1963).

%***********************************************************************
\begin{center}    {\bf ABC FLOWS } \end{center}
%***********************************************************************

\item D. McLaughlin and O. Pironneau, \label{MclPir}
{\em J. Math. Phys.}, {\bf 32} (1991) 797.  \\
 Some Notes on Periodic Beltrami Fields in Cartesian Geometry

\item P. Constatin and A. Majda, \label{ConMaj}
{\em Commun. Math. Phys.}, {\bf 115} (1988) 435. \\
   The Beltrami Spectrum for Incompressible Fluid Flows

\item Ram Ballabh, \label{Bal3}
   {\em Ganita } {\bf 1} (1950) 1. \\
On Coincidence of Vortex and Stream Lines in Ideal Liquids

%%***************************************************************************
%\begin{center}    {\bf CASIMIRS AND CURVATURE } \end{center}
%%**************************************************************************

\item A Libin and G. Sivashinsky, \label{LibSiv}
{\em Quarterly of Applied Mathematics }, {\bf 301} (1985) 1095.  \\
   Long Wavelength Instability of the ABC flows

\item Lian-Ping Wang, T. Burton, and D. E. Stock, \label{WanBurSto}
  {\em Phys. Fluids} {\bf 3A} (1991) 1073. \\
Quantification of Chaotic Dynamics for Heavy Particle
Dispersion in ABC Flow

\item Jean-Luc Gautero, \label{Gau}
{\em C. R. Acad. Sc. Paris}, {\bf 301} (1985) 1095.  \\
   Chaos Lagrangien pour Une Classe
     d' E'coulements de Beltrami

\item F. Nakamura, Y. Hattori and T. Kambe, \label{NakHatKam}
{\em J. Phys. } {\bf A25} (1992) L45. \\
Geodesics and Curvature of a Group of Diffeomorphisms
  and Motion of an Ideal Fluid

\item H. Bateman, \label{Bat}
   \underline{The Mathematical Analysis of Electrical
 and Optical Wave Motion} (Dover, 1955) see 4ff.

\item D. Struik, \label{Str}
  {\em Lectures on Classical Differential Geometry}
           (Dover, Mineola,  1988).

\item  R. Osserman, \label{Oss}
   \underline{A Survey of Minimal Surfaces}
      (Dover, New York, 1986). \\

%***********************************************************************
\begin{center}    {\bf CLEBSCH  } \end{center}
%***********************************************************************
\item A. Clebsch, \label{Cle}
{\em \"{U}ber eine allgemeine Transformation der Hydrodynamischen
 Gleichungen},
  {\em Crelle's Journal}, {\bf liv} (1859).

\item V. Zeitlin, \label{Zei}
  {\em Physica} {\bf 49D} (1991) 353.\\
Finite-mode analogs of 2D Ideal Hydrodynamics:
 Coadjoint Orbits and Local Canonical Structure

\item E. A. Kuznetsov and A. V. Mikhailov,
  {\em Physics Letters} {\bf 77A} (1980) 37.\\
On the Topological Meaning of Canonical Clebsch Variables

\item V. E. Zakharov, \label{Zak}
{\em Functional Analysis and its Applications} {\bf 23} (1989) 24. \\
The  Algebra of Integrals of Motion of Two-Dimensional Hydrodynamics
                  in Clebsch Variables

\item J. W. Herivel, \label{Her}
{\em Proceedings of the Cambridge
    Philosophical Society }, (1954) 344. \\
 The Derivation of the Equations of Motion of an ideal Fluid
        By Hamilton's Principle

\item B. Gaffet, \label{Gaf}
{\em J. Fluid Mech.} {\bf 156} (1983) 141. \\
  On Generalized Vorticity Conservation Laws

\item J. Serrin, \label{Ser}
      \underline{Mathematical Principles of
                  Classical Fluid Mechanics}
            {\em Handbuch der Physik}
          (Springer Verlag, 1977). \\
     Akron has it

\item C. Truesdell, \label{Tru}
      \underline{The Kinematics of Vorticity}
          (Indiana University Press, 1954).
 Univeristy  of Toledo has it QA 925 .T7

%***********************************************************************
\begin{center}    {\bf VARIATIONAL PRINCIPLES } \end{center}
%***********************************************************************

\item G. Mathew and M. J. Vedan, \label{MatVed}
{\em J. Math. Phys.}, {\bf 30} (1989) 949. \\
    Variational principle and Conservation Laws for
            Non-Barotropic Flows

\item H. Cendra and J. E. Marsden, \label{Mar}
{\em Physica} {\bf 27D} (1987) 63. \\
 Lin Constraints, Clebsch Potentials and Variational Principles

\item D. Holm and B. Kuperschmidt, \label{HolKup}
{\em Physica}  {\bf 6D} (1983) 347. \\
 Poisson Brackets and Clebsch Representations for Magnethydrodynamics,
         Multifluid Plasmas, and Elasticity

\item D. Holm and Y. Kimura, \label{HolKim}
{\em Phys. Fluids}  {\bf 3A} (1991) 1033. \\
 Zero Helicity Lagrangian kinematics of three dimensional advection

\item R. M. Kiehn, \label{Kie}
  {\em Continuous Topological Evolution} (1990) preprint;
  {\em Finsler Spaces, Fluctuations and the
          Topology of Space-Time} (1991) preprint

\item J. L. Schiff and W. J. Walker, \label{SchWal}
    {\em J. Math. Anal. App.} {\bf 146} (1990) 570.\\
    A Bieberbach Condition for a Class of Pseudo-Analytic
    Functions\\
See also R. J. Duffin, \label{Duf}
    {\em J. Math. Anal. App.} {\bf 146} (1990) 570.\\
Yukawan Potential Theory
\end{list}

%%%%%%%%   FIGURE CAPTIONS %%%%%%%%%%%%%%%%%%%%%%%%%%%%%%%%%%%%%%
%%%%%%%%                   %%%%%%%%%%%%%%%%%%%%%%%%%%%%%%%%%%%%%%
%%%%%%%%                   %%%%%%%%%%%%%%%%%%%%%%%%%%%%%%%%%%%%%%
\newpage
\parindent 0em
\parskip 2em
\begin{center}
{\bf FIGURE CAPTIONS}
\end{center}
\newcounter{fig}
\begin{list}%
{Fig. \arabic{fig} }{\usecounter{fig}}
\item \label{fig:cholesteric} A picture of the simplest Beltrami field
(a uniplanar Trkalian field of the first kind).
Here~$g=z$ and~$F= x +iy$. The patterns are reminiscent of the
cholesteric mesophase of liquid crystals.

\item \label{fig:helix_cos_12.1}
Here are two planes of the flow~$g=z$ and~$F= \cos(x +iy)$.
Each arrow from the lower plane to the upper plane is rotated the
same amount around the screw axis (the~$z$-axis).
In each plane, the vector field is potential flow (see Fig. 3)

\item \label{fig:helix_cos_12.2}
The bottom plane of Figure 2.

\item \label{fig:rad17_42_cos.1} A total of 714 vectors are plotted
on the surface of the sphere showing the complex Trkalian flows
for~$g=r$ and~$F= \cos(\phi +i \mbox{ arc tanh} (\cos \theta))$.
Here~$r$~is fixed to be~$0.5$. The vector magnitudes are not
drawn to scale. The true vector magnitudes are the fourth power
of those represented in the figure. This is to keep the arrows at
the poles from dominating the figure (where the vectors
become singularily large).

\item \label{fig:rad17_42_cos.2} A better look at
one of the singularities of the
complex Trkalian flow of Fig. 4.
There are a total of~$4$ singularities. Two singularities
lie at the poles (on the $z$-axis)
and two more singularities lie on opposite sides of
the sphere (at the same latitude).

\end{list}
%%%%%%%%  TABLE CAPTIONS  %%%%%%%%%%%%%%%%%%%%%%%%%%%%%%%%%%%%%%
%%%%%%%%                  %%%%%%%%%%%%%%%%%%%%%%%%%%%%%%%%%%%%%%
%%%%%%%%                  %%%%%%%%%%%%%%%%%%%%%%%%%%%%%%%%%%%%%%
\newpage
\parindent 0em
\parskip 2em
\begin{center}
{\bf TABLE CAPTIONS}
\end{center}
\newcounter{Tab}
\begin{list}%
{Table \arabic{Tab} }{\usecounter{Tab}}
\item Some elementary fields:
Note the word complex has a different connotation in its
usage in complex lamellar and complex Laplacian
(where we mean ``rescaled'' or ``conformally equivalent to)
than it does in its usage in complex Trkalian.
%That is,  complex Laplacian fields are those fields
%which have the same streamlines as (although perhaps different
%parametrizations than) Laplacian fields.
Complex Trkalian describes a vector field
whose imaginary and real parts are both Trkalian.
\end{list}%

\end{document}